\documentclass[traditabstract,twocolumn]{aa}
\usepackage{graphicx}
\usepackage{natbib}
\usepackage{amssymb}
\usepackage{hyperref}
\usepackage{txfonts}
\usepackage{rotating,bm}
\usepackage{amsmath}

\begin{document}
\title{A method to deconvolve stellar rotational velocities}
%\subtitle{Application to main--sequence field Stars}
\titlerunning{deconvolve stellar rotational velocities}
\author{
Michel Cur\'{e} \inst{1}
\and
Diego F. Rial \inst{2}
\and
Alejandra Christen \inst{3}
\and
Julia Cassetti \inst{4}
}
\institute{
Instituto de F\'{i}sica y Astronom\'{i}a, Universidad de Valpara\'{i}so, Chile
\email{michel.cure@uv.cl}
\and
Departamento de Matem\'{a}ticas, Facultad de Ciencias Exactas y Naturales, 
Universidad de Buenos Aires, Argentina \email{drial@mate.uba.ar}
\and
Instituto de Estad\'{i}stica, Pontificia Universidad Cat\'{o}lica de Valpara\'{i}so, Chile, \email{alejandra.christen@ucv.cl}
\and
Universidad Nacional de General Sarmiento, Buenos Aires, Argentina, \email{jcassett@ungs.edu.ar}
}
\date{Received December 27, 2013; Accepted March 13, 2014}
\abstract{} {Rotational speed is an important physical parameter of stars and knowing 
the distribution of stellar rotational velocities is essential for the understanding
stellar evolution. However, it cannot be measured directly but the convolution of the rotational 
speed and the sine of the inclination angle, $v \sin i$.} {We developed 
a method to deconvolve this inverse problem and obtain the cumulative distribution function (CDF) 
for stellar rotational velocities extending the work of Chandrasekhar \& M\"unch (1950)} {This 
method is applied a) to theoretical synthetic data recovering the original velocity 
distribution with very small error; b) to a sample of about 12.000 field main--sequence 
stars, corroborating that the velocity distribution function is non--Maxwellian, but is better described by 
distributions based on the concept  of maximum entropy, such as Tsallis or Kaniadakis distribution functions} {This is a very 
robust and novel method that deconvolve the rotational velocity cumulative distribution function from a 
sample of $v \sin i$ data in just one single step without needing any convergence criteria.}
\keywords{
methods: analytical -- methods: data analysis -- methods: numerical -- methods: statistical -- 
stars: fundamental--parameters -- stars: rotation
}
\maketitle

\section{Introduction\label{sec:intro}}

It is well known that all the stars rotate and the understanding about how stars rotate is essential to describe 
and modeling their formation, internal structure and evolution, how they interact with their companions, disks 
or planets. Unfortunately from observations, it is not possible to measure 
the value of their rotational velocities, but the projected value of $v \sin i$ is the measured one, where $i$ is the inclination angle. 
The standard assumption to disentangle or deconvolve the rotational velocity distribution function is assuming 
that the inclination angles are uniformly distributed over the sphere, with this assumption  
Chandrasekhar \&  M\"unch (1950) 
studied the integral equations that describe the distribution of the true  ($v$) and the apparent ($v \sin i$) 
rotational velocities, finding that the formal solution is proportional to a derivative of Abel's equation.

As Chandrasekhar \& M\"unch (1950) pointed out, the differentiation in this formal solution can lead to 
misleading results  due to an intrinsic numerical problem associated to the derivative of an Abel’s integral, 
unless the sample is of high precision. This is the main reason why this method is not usually applied. An 
alternative and  general method was introduced by Lucy (1974). 
Lucy's method is a Bayesian iterative method for deconvolve a distribution function 
assuming a prescribed formula for the kernel of eq. (\ref{eq1}), e.g., for the case of rotational velocities this kernel 
describes the projections of an uniform distribution of inclination angles $i$. 
The Lucy's method has the disadvantage that posses no convergence 
criteria and the requested number of iterations is only justified {\it{a posteriori}} in view of the results. 
Nevertheless, the Lucy's method is widely used in the astronomical community to disentangle distribution 
functions from different observations samples.

In this article we enhanced the pioneer work of  Chandrasekhar \& M\"unch (1950), we integrate the probability 
distribution function (PDF) for the rotational velocities and obtain the cumulative distribution function (CDF) for 
the velocities. This CDF is attained in just {\it{one step}}, without the need of a convergence criteria usually 
necessary in iterative methods, giving robustness to our novel method.

This article is structured as follows: In Section 2  we present the mathematical description of the method. In 
section 3 we apply it to a theoretical example of $v \sin i$, we show that $v \sin i$ is given by a $\chi$ 
distribution when  the velocity distribution comes from a Maxwellian distribution (Deutsch 1970).  Our method recovers the 
Maxwellian distribution with a very high degree of confidence. We also discuss in this section the relation 
between the length of the sample and the error in the CDF. Section 4 is advocated to a real sample of circa 
12000 main--sequence field stars. We obtain for the first time the true CDF for the velocities of this sample.
Furthermore,  we recover the results of Carvahlo et al. (2009) demonstrating that the rotational velocity 
distribution function for this sample is non--Maxwellian, but is better described by a Tsallis or Kaniadakis 
distribution functions. In the last section we discuss our conclusions and future work.

\section{The Method\label{sec:method}}

An important class of inverse problems in astronomy has the following form:
\begin{equation}
\label{eq1}
f_{Y}(y)=\int p(y\,|\,x)\,f_{X}(x)dx,
\end{equation}
which is known as Fredholm integral problem of the first kind (Lucy 1994), where $f_{Y}$ is a function accessible 
to observation and $f_{X}$ is the function of interest. The kernel $p$ of this integral is related to the remoteness 
of the measurement process.\\
Different approaches are considered in the literature  (see Lucy 1994 and references therein), almost all of 
them based on maximum likelihood, where the optimization is carried out using Bayesian-based iterative 
methods (see Richardson 1972).

In many problems, the function of interest is the cumulative distribution $F_{X}$ of the random variable  $X$  
and different approaches could be applied. For a special form of the kernel $p$ 
(related with the applications) we can invert the integral problem (\ref{eq1}). 

\subsection{The Basic Problem}

The problem which we wish to consider may be formulated in the following form:
a positive continuous random variable $X$ occurs with a probability given by a unknown density function $f_{X}$.
The probability density function $f_{Y}$ of a observed variable $Y$ is related with $f_{X}$ by equation (\ref{eq1}),
where $p$ is the kernel or conditional probability density function, i.e.:
\begin{equation}
p(y\,|\,x) \,dy=P(Y\in [y,y+dy)\,|\,X=x).
\end{equation}
The mathematical problem consists in obtaining $f_{X}$ from the theoretical function $p$ and the observed distribution $f_{Y}$.
We study the particular case where $Y$ is a projection of $X$: $p(y\,|\,x)=0$ if $y>x$ or $y<0$ and depends 
only on the ratio $s=y/x$.
In this case we can write $p(y\,|\,x)\,dy=q(y/x)\,dy/x$ and then
\begin{equation}
\label{basic}
F_{Y}(y)=1-\int_{y}^{\infty}Q(y/x)\,f_{X}(x)\,dx,
\end{equation}
where $F_{Y}$ the cumulative distribution function.

Chandrasekhar \& M\"unch (1950) considered the integral equation governing the distribution of the true an the 
apparent rotational velocities of stars, $y=x\sin i$, where $x=v$ and $i$ is the inclination angle, assuming an 
uniform distribution 
over the sphere (a detailed derivation of $Q(y/x)$ is given in Appendix \ref{AppendixA}).  In this case, the integral 
equation reads as follows:
\begin{equation}
\label{eq-problema}
f_{Y}(y)=\int_{y}^{\infty}\dfrac{y}{x\sqrt{x^{2}-y^{2}}}\,f_{X}(x)dx.
\end{equation}
They obtain the solution of this problem, based on the formal solution of Abel's integral equation, namely:
\begin{equation}
\label{eq-f-metodo}
f_{X}(x)=-\dfrac{2}{\pi}x^{2}\dfrac{\partial}{\partial x}x\int_{x}^{\infty}\dfrac{1}{y^{2}\sqrt{y^{2}-x^{2}}}\,f_{Y}(y)dy.
\end{equation}
which is not of much practical use, since it requires differentiation of a functional of the observed density function $f_{Y}$ 
and it can lead to wrong results (see Chandrasekhar \& M\"unch 1950).

From the definition of CDF of a random variable:
\begin{equation*}
F_{X}(x)=\int_{0}^{x}f_{X}(\xi)d\xi=1-\int_{x}^{\infty}f_{X}(\xi)d\xi
\end{equation*}
and using Equation (\ref{eq-f-metodo}), we obtain:
\begin{equation*}
F_{X}(x)=1+\dfrac{2}{\pi}\int_{x}^{\infty}\xi^{2}g'(\xi)\,d\xi,
\end{equation*}
where $g(\xi)=\xi\displaystyle{\int_{\xi}^{\infty}\dfrac{1}{y^{2}\sqrt{y^{2}-\xi^{2}}}\,f_{Y}(y)\,dy}$.
After applying integration by parts, we get:
\begin{equation}
\label{def-cdf}
F_{X}(x)=\,1+\left.\dfrac{2}{\pi}\,\xi^{2}g(\xi)\right|_{\xi=x}^{\xi=\infty}
-\dfrac{4}{\pi}\int_{x}^{\infty}\xi\, g(\xi)\,d\xi.
\end{equation}
Using the following inequality 
\begin{align*}
\xi^{2}g(\xi)=&\,\int_{\xi}^{\infty}\frac{\xi^{3}}{y^{2}\sqrt{y^{2}-\xi^{2}}}f_{Y}(y)\,dy\\
\leq&\,\sup_{y\geq\xi}f_{Y}(y)\int_{\xi}^{\infty}\frac{\xi^{3}}{y^{2}\sqrt{y^{2}-\xi^{2}}}\,dy\\
=\,&\xi\sup_{y\geq\xi}f_{Y}(y)\leq \sup_{y\geq\xi}y f_{Y}(y),
\end{align*}
it holds that $\lim\limits_{\xi\to\infty}\xi^{2}\,g(\xi)=0$, provided $\lim\limits_{y\to\infty}y\,f_{Y}(y)=0$. 
The latter assumption is true for all the known distribution functions. Therefore, re--arranging Eq. (\ref{def-cdf}), we get:
\begin{align}
\label{eq: inv}
F_{X}(x)=&\,1-\dfrac{2}{\pi}\,\int_{x}^{\infty}\dfrac{x^{3}}{y^{2}\sqrt{y^{2}-x^{2}}}f_{Y}(y)dy
-\dfrac{4}{\pi}\int_{x}^{\infty}\xi\, g(\xi)\,d\xi.
\end{align}
Interchanging order of integration, the last integral can be written as
\begin{align*}
\int_{x}^{\infty}\xi\, g(\xi)\,d\xi=&\,
\int_{x}^{\infty}\int_{\xi}^{\infty}\dfrac{\xi^{2}}{y^{2}\sqrt{y^{2}-\xi^{2}}}f_{Y}(y)dy\,d\xi\\
=&\,\int_{x}^{\infty}\left(\int_{x}^{y}\dfrac{\xi^{2}}{y^{2}\sqrt{y^{2}-\xi^{2}}}\,d\xi\right) f_{Y}(y)dy\\
=&\,\frac{1}{2} \int_{x}^{\infty}\left(\frac{x \sqrt{y^2-x^2}}{y^2}+\arccos(x/y)\right) f_{Y}(y)dy.
\end{align*}
and replacing in \eqref{eq: inv}, finally  we obtain:
\begin{equation}
\label{eq-F-metodo}
F_{X}(x)=\,1-\dfrac{2}{\pi}\,\int_{x}^{\infty}
\left(\frac{x}{\sqrt{y^2-x^2}}+\arccos(x/y)\right) f_{Y}(y)\,dy.
\end{equation}

This equation provides a numerically stable method for solving (\ref{eq1}) in this particular case. 
Our purpose is to develop a novel algorithm for solving the general problem (\ref{eq1}), like an 
alternative to the iterative method proposed by Lucy (1974) and recover the original 
Chandrasekhar \& M\"unch
method, without introducing numerical instabilities due to derivative (see equation \ref{eq-f-metodo}).

\section{A Theoretical Test}

\begin{figure*}
 \centering
 \includegraphics{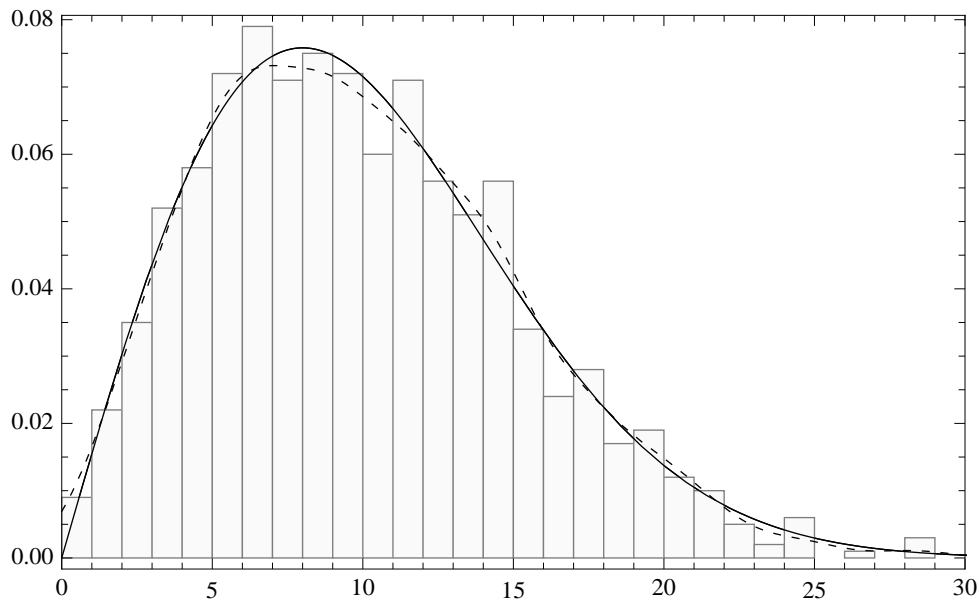}
 \caption{Histogram of a synthetic sample of $y=v \sin i$ with $n=1000$ from the PDF (eq. \ref{chi2ysig}), with $\sigma=8$. 
 In solid black line is plotted the corresponding theoretical (Maxwellian) PDF. Dashed--line shows the corresponding KDE of 
 this sample calculated with an Gaussian kernel.}
 \label{Fig2}
\end{figure*}

In this section we evaluate the proposed method in section \ref{sec:method}. In order to do that, we are going to assume that the distribution of velocities is given by a Maxwellian distribution (as 
it is considered by Deutsch 1970) with dispersion $\sigma$, i.e., 

\begin{equation}
\label{PDF-Maxwell}
f_{X}(x)=\sqrt{\frac{2}{\pi }} \frac{ x^2 }{\sigma ^3} e^{-\frac{x^2}{2 \sigma ^2}}
\end{equation}
where $x>0$. The behavior of this distribution is shown in Figure \ref{Fig2} in solid line.

Therefore, eq. (\ref{eq-problema}) gives:
\begin{equation}
%f_{Y}(y)= \int_{y}^{\infty} \left(\frac{y}{x \sqrt{x^2-y^2}}\right) \,   \left(\sqrt{\frac{2}{\pi }}\frac{ x^2 }{\sigma ^3} 
f_{Y}(y)= \int_{y}^{\infty} \frac{y}{x \sqrt{x^2-y^2}} \,  \sqrt{\frac{2}{\pi }}\frac{ x^2 }{\sigma ^3} e^{-\frac{x^2}{2 \sigma ^2}} dx
\end{equation}
This integral has an explicit analytic solution which is:
\begin{equation}
\label{chipar2}
f_{y}(y)=\frac{y }{\sigma ^2} \,e^{-\frac{y^2}{2 \sigma ^2}}
\end{equation}
for $y>0$. Considering that the $\chi$--distribution is defined by:
\begin{equation}
f_{\chi}(\nu;z)=\frac{2^{1-\frac{\nu }{2}}
   e^{-\frac{z^2}{2}} \, z^{\nu
   -1}}{\Gamma \left(\frac{\nu
   }{2}\right)}
\end{equation}
where $\nu$ is a real positive parameter, $z$ is any positive real number and $\Gamma$ is the Gamma function. 
Then, equation (\ref{chipar2}) correspond to:
\begin{equation}
\label{chi2ysig}
f_{Y}(y)= \frac{1}{\sigma}\,f_{\chi}(2;\frac{y}{\sigma}).
\end{equation}

In order to apply this method, we first create a sample of $v \sin i$ values using equation (\ref{chi2ysig}), 
i.e., generating $n$ random values $\left\{ Y_1,Y_2,\cdots, Y_{n-1},Y_n\right\} $, then we obtain $\hat{f}_Y(y)$
using a Kernel Density Estimator method (KDE) for this sample % in order to estimate $f_Y(y)$ 
and insert it in equation (\ref{eq-F-metodo}).
The KDE is a non--parametric method to estimate the probability distribution of a random variable 
(in our case  $y=x \sin i$) from a random sample.

The KDE estimator is defined as follows:
\begin{equation}
\hat{f}_{Y}(y)=\frac{1}{n\, h} \sum_{j=1}^{n} K\left(  \frac{y-Y_j}{h} \right),
\end{equation}
where $K$ is the kernel function and $h$ is the bandwidth. We use in this work a {\it{Gaussian}} kernel 
$K_G(y)=1/\sqrt{2\pi} \, e^{-\frac{1}{2} y^2} $, because this kernel smooths the 
distribution.
 The bandwidth is defined as: $$h=0.9 \min\left( \tilde{\sigma}, \frac{IQ}{1.34}\right) \,n^{-1/5}\,\,,$$  where $\tilde{\sigma}$ is the standard deviation 
of the random variable under study, $Y$, and $IQ$ is the correspondent interquartile range  (Silverman 1986).
Figure \ref{Fig2} shows an histogram of a synthetic sample of $n=1000$, 
and the corresponding KDE function is shown in dashed line.

The steps of our algorithm to get the estimated CDF are the following:
i) Have a sample of $v \sin i$, ii) Obtain a KDE from this sample using a suitable kernel 
function, and iii) Calculate the estimated CDF using equation (\ref{eq-F-metodo}) 
with KDE (from previous step) as the $f_Y(y)$ function.

We have to point out here, that exists an  {\it{arbitrary choice}} in the number of values of 
$x$ that we get from equation (\ref{eq-F-metodo}). A reasonable restriction is to take a $\Delta x$ not lower than 
the sample's measurement error. For the sample we use in section (\ref{sec4}) the error is $\Delta x=1\, km/s$, 
value that we use in this theoretical example.

\begin{figure*}
 \centering
 \includegraphics{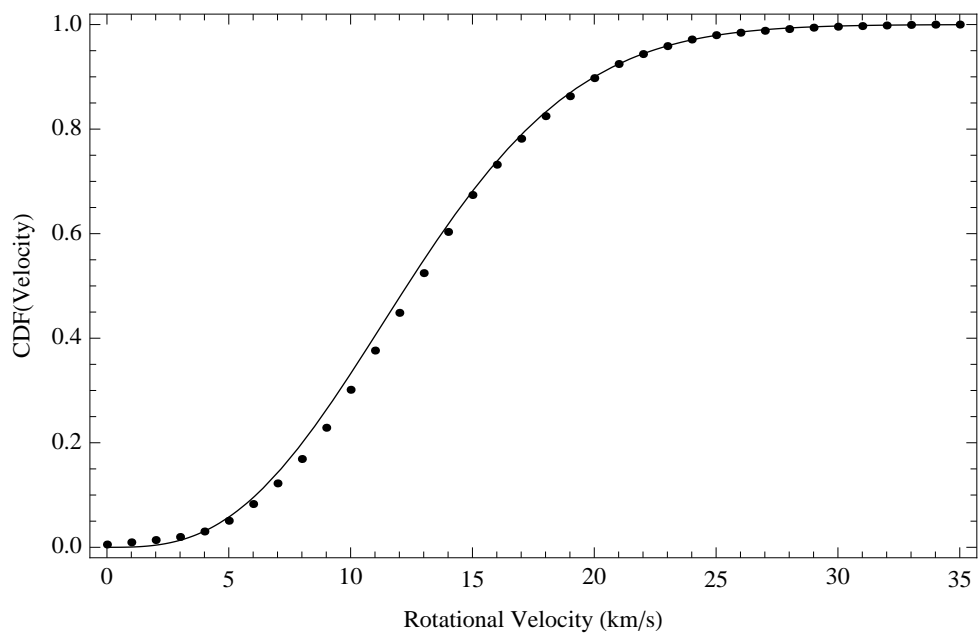}
 \caption{The CDF values calculated  from eq. (\ref{eq-F-metodo}) with a step of 1 $km/s$ are plotted in black dots. The theoretical Maxwellian distribution is shown in black solid line. }
 \label{Fig3}
\end{figure*}

Figure (\ref{Fig3}) shows the estimated CDF in the interval 0 to 35 $km/s$, with a step of  1 $km/s$, for a 
synthetic sample of $v \sin i$ of size 
$n=1000$ and parameter $\sigma=8$.  These data were obtained directly using our method described 
previously, without needing any convergence 
criteria. From this empirical CDF we can calculate the moments of the distribution (e.g., mean, variance), 
but the approach for the moments from Chandrasekhar \& M\"unch (1950)
is more straightforward. Nevertheless, with this new method we can calculate percentiles, intervals and in 
general any probability associated to the velocity distribution.

Figure (\ref{Fig3}) also shows the theoretical CDF (solid line) of this Maxwellian distribution for the same parameter, i.e.,
\begin{equation}
F_M(x)=\text{erf}\left(\frac{x}{\sqrt{2} \sigma
   }\right)-\sqrt{\frac{2}{\pi }} \frac{x} {\sigma } e^{-\frac{x^2}{2 \sigma^2}}  ,
\label{CDF-Maxwell}
\end{equation} 
where $\text{erf}$ corresponds to the error function.

In order to get an estimation of our method's error, we use the standard discrete Integrated Square Error (ISE) defined as:
\begin{equation}
\mathrm{ISE}=\frac{1}{n}  \sum_{j=1}^{n} \left(CDF(x_{j})-F_M(x_j)     \right)^2
\end{equation}
and afterwards we obtain the standard discrete Mean Integrated Square Error (MISE) by computing the mean of ISE 
for several samples of a given fixed size $n$. %We create two different sets with sample sizes $n=1000$ and $n=10000$.
Therefore, we run a Monte Carlo simulation calculating  ISE for a set 
of 500 samples of $n=1000$ data and another set of 500 samples with 
$n=10000$ data. Figure (\ref{Fig5}) shows the histogram for 
both sets of samples, and Table (\ref{table:1}) 
shows the Mean and Median values of our simulations, where the Mean value of each set represents
the estimated MISE  of our model. Clearly the more 
data has the sample, the lower is the error in the 
estimation of the CDF.

\begin{figure*}
 \centering
 \includegraphics{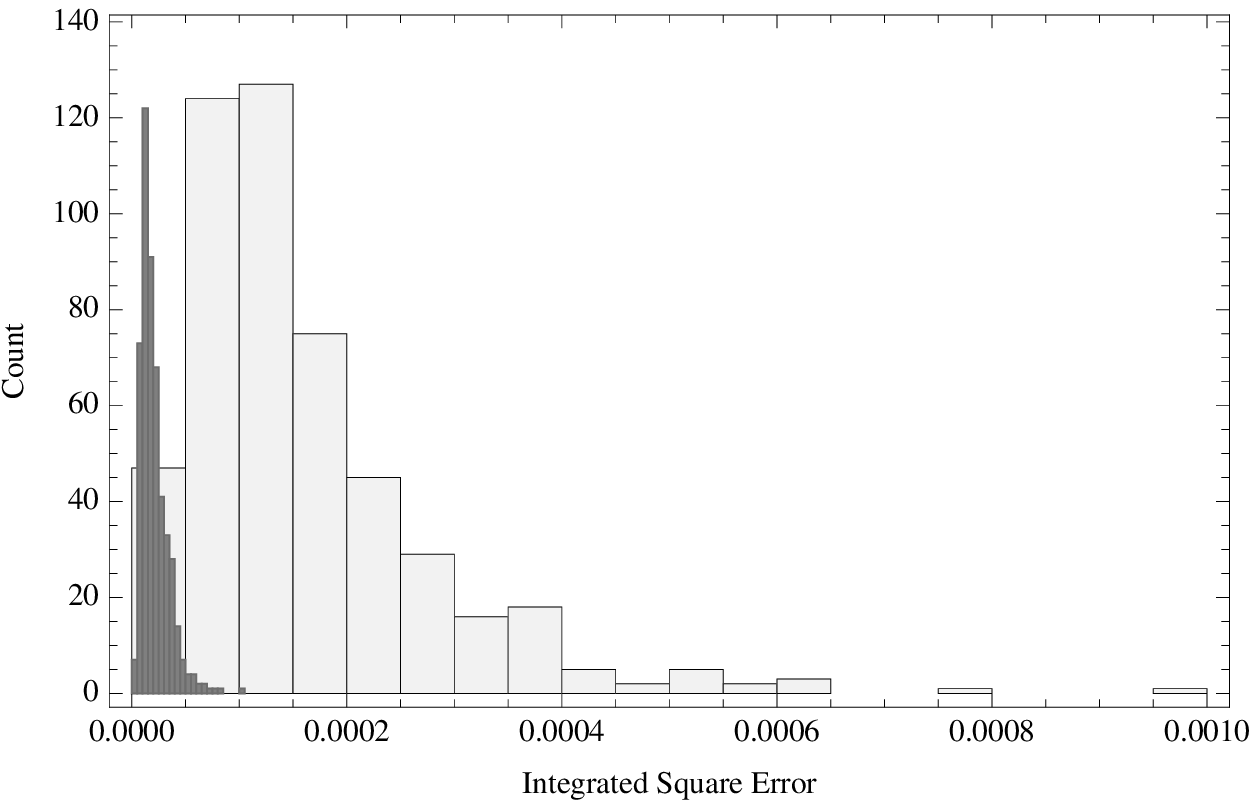}
 \caption{Histogram of the ISE for a set of 500 samples with 1000 data each, shown in light--gray and a 
 set of 500 samples of 10000 data each shown in dark--gray (see text for details).}
 \label{Fig5}
\end{figure*}
\begin{table}
\caption{MISE}           
\label{table:1}   
\centering                 
\begin{tabular}{r c c }       
\hline\hline                 
Sample & Mean & Median \\    
\hline                        
1000 	&  $1.60 \times 10^{-4}$  & $1.28 \times 10^{-4}$ \\
10000 	&  $2.09 \times 10^{-5}$ &  $1.71\times  10^{-5}$\\
\hline
\end{tabular}
\end{table}

The MISE depends mostly on the  KDE of each sample used for computation of the  
CDF, when the KDE from the sample is improved, so does the estimation of CDF. \\
We can conclude that our method 
recovers the Maxwellian distribution with very high precision.

\section{Application to Main--Sequence Field Stars \label{sec4}}

In this section we apply this method to a large sample of 
measured $v \sin i$ data. We use the sample from the  Geneva--Copenhagen survey of the 
Solar neighborhood (Nordstr\"om et al. 2004 and Holmberg et al. 2007), which contain information about 
$16500$ F and G main--sequence field stars. In this catalog $12931$ stars have values of  $v \sin i$. Furthermore, 
considering that the values of $v \sin i$ are given rounded to the nearest $km/s$ up to $30 km/s$, upwards that 
value the sample gives $v \sin i$  to the nearest $5$ or $10$ $km/s$ and that about $91\%$ of the sample have 
rotational  speeds below  $30 km/s$,  we select (following Carvalho et al. 2009) only  stars with 
$v \sin i  \leq 30 km/s$, giving a  
sample of $11818$ stars.

\begin{figure*}
 \centering
 \includegraphics{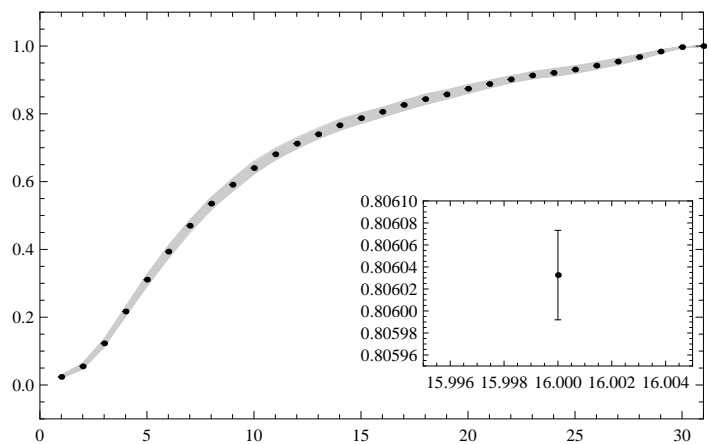}
 \caption{The estimated CDF from the sample of $11818$ stars are shown in 
 black dots, together with an approximated confidence interval (error bar).
 In light--gray all the bootstrap--CDFs are also plotted. The Zoomed plot shows the error bar calculated using the methodology from Efron (1993). }
 \label{Fig4}
\end{figure*}

Applying our method  we obtained the estimated CDF from this sample, as shown in figure \ref{Fig4} 
in black dots. In order to estimate the error of the estimated CDF  we create $1000$ bootstrap samples of the original one and calculate the corresponding  $1000$ bootstrap--CDFs. In
Figure \ref{Fig4} we also plotted in light--gray all these bootstrap--CDFs. Concerning the error's estimator 
we add in each point of Figure \ref{Fig4} an error bar which is the $95\%$ confidence interval  (see the zoomed plot in Figure \ref{Fig4}). Since the estimated bootstrap--CDFs from the original sample are functions, it is difficult to get the distribution of probability of the estimator. Therefore in order to calculate the probability of the original estimated CDF, we used the approximated confidence interval with normal standard percentile $Z_{\alpha/2}$, namely:
\begin{equation*}
\widehat{CDF} \pm Z_{\alpha/2} \widehat{se}\, \, ,
\end{equation*}
where $\widehat{se}$ is the standard deviation of Bootstrap samples, and $Z_{\alpha/2}$ is the value in which the standard normal distribution accumulate $97.5\%$ of the area under its PDF and $\alpha=0.05(=1-0.95)$ is the complement of the confidence. This procedure is the simplest one for calculating confidence intervals from Bootstrap method proposed in Efron (1993).

One of the advantages of getting the CDF is to give information about the rotational velocity distribution. Observing 
the black dots from  figure \ref{Fig4}, we see that approximately  $50\%$ from sample have a magnitude less than  
$7 \, km/s$ and  $10\%$ of the fastest rotational speeds are approximately between $24 \, km/s$ and $30 \, km/s$.

Deutsch (1970) proved, using methods of classical statistical mechanics, that the distribution of rotational speeds
is given by a Maxwellian Distribution (eq. \ref{PDF-Maxwell}), under the assumption that the angles are distributed 
uniformly over the sphere. Therefore, in order to describe the rotational velocity distribution function, we fit a 
Maxwellian CDF  to our estimated CDF. The fitting method minimizes the Least Squares, i.e., 
$\phi= \frac{1}{n} \sum_{j=1}^{n} \left(Y_j-F_M(X_j) \right)^2$,
where $F_M$ is the CDF for the Maxwellian Distribution (eq. \ref{CDF-Maxwell}). For the minimization process we 
use the Nelder--Mead simplex algorithm (Nelder \& Mead 1965) .\\

Figure \ref{Fig5} shows the estimated CDF and also shows the Maxwellian's CDF (black solid line), the minimization gives a dispersion  of $\sigma=5.64$ that corresponds for the minimum value of $\phi=0.0087$. Clearly from this figure, a Maxwellian 
distribution cannot describe the proper distribution of the rotational speeds of this sample.
Indeed, the estimated CDF has tails with increased positive probability than those of Maxwellian's CDF, i.e., 
the distribution of rotational velocity has more dispersion than the one from a Maxwellian distribution.
\begin{figure*}
 \centering
 \includegraphics{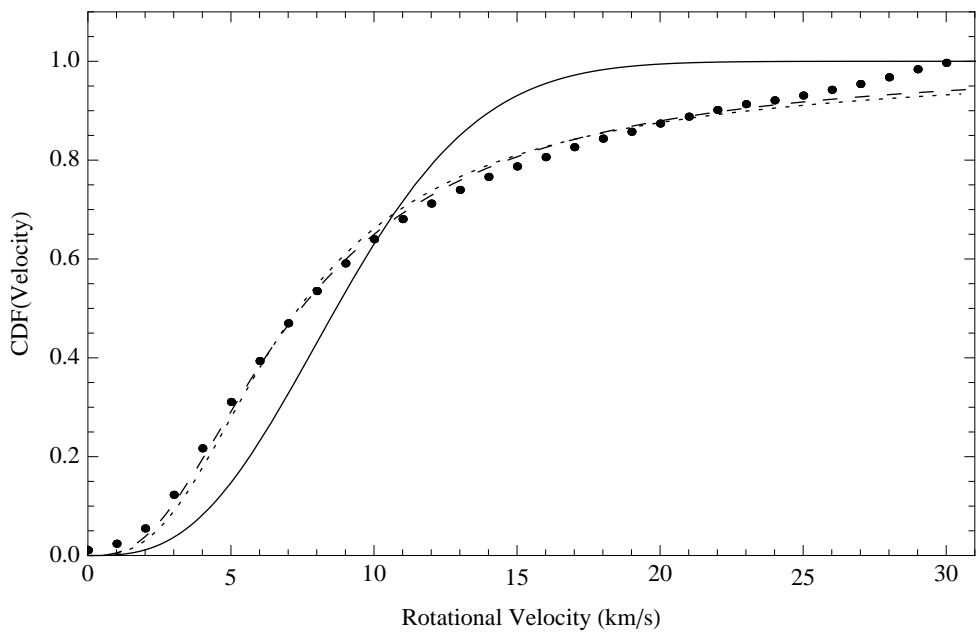}
 \caption{The empirical CDF from the sample in black dots without error bar. 
 The fitted CDF function from a Maxwellian distribution with $\sigma=5.64$ is shown in 
 black solid line, fitted Tsallis distribution with $\sigma_q=3.46$ and $q=1.41$ is shown 
 in dashed line and fitted Kaniadakis distribution with $\sigma_k= 4.64$ and $k=0.45$ is shown in dotted line.}
 \label{Fig5}
\end{figure*}

This sample has been used by Carvalho et al. (2009), who performed a statistical study of it, showing that the
empirical distribution function of this sample cannot be fitted using a {\it{Maxwellian}} distribution 
(Deutsch 1970). They obtained a much better fit when the assumption of  {\it{Gibbs entropy}} in standard 
statistical mechanics is released and  distribution functions from non--extensive statistical mechanics are applied 
to this sample. Specifically they used the {\it{Tsallis}} distribution (Tsallis 1998) and the {\it{Kaniadakis}}  power--law 
distribution (Kaniadakis 2002, 2005), both based on the concept of {\it{maximum entropy}} (Gell-Mann \& Tsallis 2004).

In figure \ref{Fig5} these distributions are also plotted. We can see that for left tail, the estimated CDF (black dots) 
and both Tsallis and Kaniadakis distributions have velocity values below $8 \, km/s$ representing the $60\% $ of the 
data, while for the Maxwellian distribution this percentage is reached about $10\, km/s$. Concerning the right tail, the Maxwellian 
distribution for the rotational velocity less than $17\, km/s$ has probability  about $95\%$ while for the other two distributions
this probability is reached for rotational speeds with values above $25\,km/s$.

Having a known integral expression (given by eq. \ref{eq-F-metodo}), our method allows to fit any known distribution. 
In this case, we arise to the same conclusion that Carvalho et al. (2009), but using a  different algorithm, i.e., 
Tsallis or Kaniadakis distributions are in a close agreement to the empirical CDF, although both distributions show a 
slightly discrepancy in the tails.

\begin{table}
\caption{Nonlinear Model Results}             % title of Table
\label{table:2}      % is used to refer this table in the text
\centering                          % used for centering table
\begin{tabular}{l c c c}        % centered columns (4 columns)
\hline\hline                 % inserts double horizontal lines
Distribution & $LS$ & $\sigma$ & Second parameter \\    % table heading 
\hline                        % inserts single horizontal line
Maxwellian 	& 0.3142 & 5.639 & \dots \\      % inserting body of the table
Tsallis 			& 0.0126 & 3.461 & $q=$1.413 \\
Kaniadakis 	& 0.0217 & 4.638 & $k=$0.445 \\
\hline                                   %inserts single line
\end{tabular}
\end{table}

\section{Discussion and Conclusions}

In this work we have obtained  the cumulative distribution function of ''de-projected'' velocities.
It is well known that from the estimated CDF we can obtain probabilities in general, e.g., the distributional moments, 
the probability of an interval, median, percentiles and any other statistical feature of the sample.

This novel presented method is an extension of  the pioneer work introduced by Chandrasekar 
and M\"unch (1950) allowing to obtain the CDF without numerical instability caused by the use of derivative, 
that was the main disadvantage of the PDF from Chandrasekar and M\"unch (1950). 
Furthermore, this estimated CDF is obtained in just one step without needing any convergence criteria 
in comparison with the widely used  iterative method of Lucy (1974).\\

Deutsch (1970) shows that if the direction of the rotational velocity is uniformly distributed and each Cartesian
component is distributed independently of the other ones, then the magnitude of
the velocity follows a Maxwellian distribution law. However, the independence assumption is not clear and 
we proved in the previous section that the Maxwellian distribution does not fit accurately the empirical CDF.

On the other hand, Carvalho et al. (2009) found a better agreement for two other 
probability distributions,namely: Tsallis and Kaniadakis. These distributions are based on the concept  
of maximum entropy. In our case, using the new method, we confirm their analysis but this time 
from the rotational velocity estimated CDF. This results open the question about the validity of the assumption 
of Maxwellian distribution.

 {\it{Future work: }} We will extend the applicability of our model for samples that 
show  bi--modal velocity distributions, e.g., the 
data from VLT FLames Tarantula Survey (Ram\'irez-Agudelo et al, 2013).

There exits samples where the number of stars with very low rotational velocities 
are a few or none at all (see e.g., Yudin 2011, Ram\'irez-Agudelo et al, 2013). If the distribution of 
inclination angles is uniform, then independently of the rotational velocity distribution, there must be a 
non--negligible portion of the sample with lower values of the projected rotational velocity due to low values of $i$. 
There are some studies of open clusters where this assumption seems no longer reasonable, see Silva et al. (2013) and 
Rees y Zijlstra (2013).  We want to develop a "completeness" test for the uniform distribution of angles of  a sample
of $v \sin i$.

Furthermore, we want to use a general function in order to describe an arbitrary orientation of rotational axes and 
study the distribution of rotational velocities in a more general description. Finally a very ambitious project will be to obtain {\it{simultaneously}} both distributions from a data sample.

\begin{acknowledgements}
This work has been partially supported by project CONICT-Redes 12-0007.
MC thanks the support of FONDECYT project 1130173 and Centro de Astrof\'isica de Valpara\'iso. 
JC thanks the financial support from project: "Ecuaciones Diferenciales y An\'alisis Num\'erico" - Instituto
de Ciencias, Instituto de Desarrollo Humano e Instituto de Industria - Universidad Nacional
de General Sarmiento. DR acknowledge the support of project PIP11420090100165, CONICET.
AC thanks the support from Instituto de Estad\'{i}stica, Pontificia Universidad Cat\'{o}lica de Valpara\'{i}so.
\end{acknowledgements}

\begin{appendix} 
\section{Distribution of projected angles}
\label{AppendixA}
\begin{figure}[h!]
\label{esfera}
\begin{center}
\includegraphics[scale=0.8]{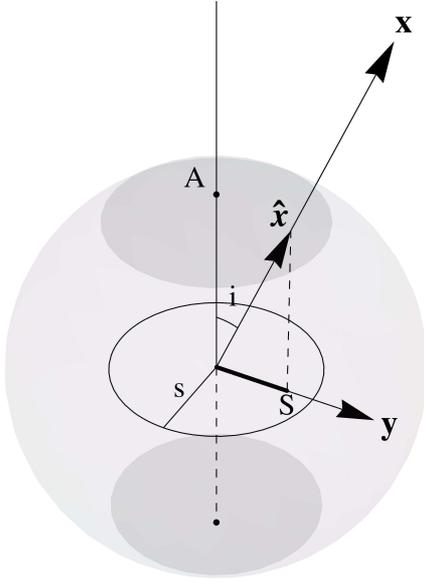}
\caption{Distribution of $\sin i$}
\end{center}
\end{figure}
Let $\mathbf{x}$ be a random vector in $3$-D and $\mathbf{y}$ be the projection of $\mathbf{x}$
to the plane normal to the line of sight. If $X=\|\mathbf{x}\|$, $Y=\|\mathbf{y}\|$, $i$ is
the angle between $\mathbf{x}$ and the line of sight, and $S=|\sin(i)|=Y/X\in [0,1]$,
the condition $S\leq s$ is equivalent to $\hat{\mathbf{x}}=\vec{\mathbf{x}}/X\in A$,
where $A$ is the {\it polar region} of the unit sphere corresponding to the inclination angle
$i\in [0,i_{0}]\cup [\pi-i_{0},\pi]$,
azimuthal angle $\phi\in [0,2\pi]$ and $i_{0}=\arcsin(s)$ (see figure \ref{esfera}).
Asume $X$, $\hat{\mathbf{x}}$ are independent and $\hat{\mathbf{x}}$
is uniformly distributed over the unit sphere, i. e. $P(\hat{\mathbf{x}}\in A)$ is proportional to the area of $A$.
Therefore
\begin{align*}
P(S\leq s)=&\,\frac{\textrm{\text{Area}}(A)}{4\pi}
=\frac{1}{2\pi}\int_{0}^{i_{0}}\int_{0}^{2\pi}\sin(i)\,d\phi\,di\\
=&\,\int_{0}^{i_{0}}\sin(i)\,di
=1-\cos(\arcsin(s))=1-\sqrt{1-s^{2}}.
\end{align*}
Since the PDF verifies $f_S(s)=d\,P(S\leq s)\, /ds$, we have $f_{S}(s)=s/\sqrt{1-s^{2}}$, and replacing in \eqref{eq1} we get
\begin{align}
\label{eq: int direct}
f_{Y}(y)=\int_{y}^{\infty}\dfrac{y}{x\sqrt{x^{2}-y^{2}}}\,f_{X}(x)\,dx.
\end{align}
which corresponds to eq. [9] from Chandrasekhar \& M\"unch (1950).

\section{Tsallis \& Kaniadakis Distributions}
A standard assumption of statistical mechanics, based on the {\it{Gibbs entropy}} is that quantities, such as the  energy 
are {\it{extensive}} variables, i.e., the total energy of the system is proportional to the system size; similarly the 
entropy is also supposed to be extensive. Tsallis  statistical mechanics is based on the concept  of maximum 
entropy (see details in Gell-Mann \& Tsallis 2004).
From a Mathematical point of view, Tsallis PDF is defined as:
\begin{equation}
f_q(x)=\left(\frac{4 (q-1)^{3/2} \Gamma \left(\frac{1}{q-1}\right)}{\sqrt{\pi } \sigma_q ^3 \Gamma
   \left(\frac{1}{q-1}-\frac{3}{2}\right)} \right) x^2 \left(1-\frac{(1-q) x^2}{\sigma_q ^2}\right)^{\frac{1}{1-q}}
\end{equation} 
where $\sigma_q$ is the dispersion. When the $q$--parameter tend to $1$ the standard Maxwellian distribution is attained.
The CDF of this distribution is:
%\begin{eqnarray}
%F_q(x)&=& \left(  2 \sqrt{\pi } \sqrt{q-1} \sigma_q ^3 \Gamma \left(\frac{1}{2}+\frac{1}%{q-1}\right) \left((q-1) x^2+\sigma_q ^2\right)  \right)^{-1}  \left(\sigma_q ^{\frac{2}{q-1}-3} x^{\frac{q+3}{1-q}} \Gamma \left(\frac{1}{q-1}\right) \left(-q-\frac{\sigma_q
%   ^2}{x^2}+1\right)^{\frac{1}{1-q}} \right)  \nonumber \\
%& & \times \, \left[ e^{\frac{i \pi }{q-1}} \left((9 q-17) \sigma_q ^4 x^{\frac{2 q}{q-1}}+x^{\frac{2}{q-1}} \left(-(q-3)^2 (q-1) x^6+(q
%   (7 q-18)+7) \sigma_q ^2 x^4+\sigma_q ^6\right)\right)   \right. \nonumber \\
%& & \left. + \,\sigma_q ^{-\frac{2}{q-1}} x^{-\frac{2}{q-1}} \left(\frac{(q-1) x^2}{\sigma_q ^2}+1\right)^{\frac{1}{1-q}}
%   \left(-q-\frac{\sigma_q ^2}{x^2}+1\right)^{\frac{1}{q-1}} \left(\sigma_q ^2 x^{\frac{2}{q-1}}+(q-1) x^{\frac{2
%   q}{q-1}}\right)  \times \right. \nonumber \\
%& & \left. \left( 2 \sigma_q ^2 x^{\frac{4}{q-1}} \left((q-1) x^2+\sigma_q ^2\right) \,
%_2F_1\left(1,\frac{1}{2}+\frac{1}{1-q};\frac{1}{2};-\frac{(q-1) x^2}{\sigma_q ^2}\right) \right. \right.\nonumber \\
%& & \left. \left. +(q-3) (3 q-5) x^{\frac{4
%   q}{q-1}}-\sigma_q ^2 x^{\frac{4}{q-1}} \left(4 (2 q-3) x^2+3 \sigma_q ^2\right) \right) \right]
%\end{eqnarray}
\begin{eqnarray}
F_q(x)&=& \left(  2 \sqrt{\pi } \sqrt{q-1} \sigma_q ^3 \Gamma\left(\frac{1}{2}+\frac{1}{q-1}\right) \left((q-1) x^2+\sigma_q ^2\right)\right)^{-1} \,
\sigma_q ^{\frac{2}{q-1}} x^{\frac{q+3}{1-q}} \Gamma \left(\frac{1}{q-1}\right) \left(-q-\frac{\sigma_q ^2}{x^2}+1\right)^{\frac{1}{1-q}} \nonumber \\
& & 
  \left[ \sigma_q ^{-\frac{2}{q-1}} x^{-\frac{2}{q-1}} \left(\frac{(q-1) x^2}{\sigma_q ^2}+1\right)^{\frac{1}{1-q}} \left(\sigma_q ^2
   x^{\frac{2}{q-1}}+(q-1) x^{\frac{2 q}{q-1}}\right) \left(-q-\frac{\sigma_q ^2}{x^2}+1\right)^{-\frac{1}{1-q}} \right. \nonumber \\
 & &  
\left.    \left(2 \sigma_q ^2 x^{\frac{4}{q-1}} \left((q-1) x^2+\sigma_q ^2\right) \, _2F_1\left(1,\frac{1}{2}+\frac{1}{1-q};\frac{1}{2};-\frac{(q-1) x^2}{\sigma_q
   ^2}\right)   \right. \right.\nonumber \\
& & \left. \left.  
+(q-3) (3 q-5) x^{\frac{4 q}{q-1}}-\sigma_q ^2 x^{\frac{4}{q-1}} \left(4 (2 q-3) x^2+3 \sigma_q ^2\right)\right) \right. \nonumber \\
& & \left.  
+e^{\frac{i \pi }{q-1}} \left((9 q-17) \sigma_q ^4 x^{\frac{2 q}{q-1}}+x^{\frac{2}{q-1}} \left(-(q-3)^2 (q-1) x^6+ 
(q (7 q-18)+7) \sigma_q ^2 x^4+\sigma_q^6\right) \right) \right] 
\end{eqnarray}

here $_2 F_1$ is the Gauss Hypergeometric function and $\Gamma$ is the  Gamma function.\\

Kaniadakis (2002, 2005) based on the same concept of maximum entropy, developed a power-law statistics, where the distribution function is given by:
\begin{equation}
f_k(x)= \frac{8 \sqrt{\frac{2}{\pi }} k^4 \Gamma \left(\frac{7}{4}+\frac{1}{2 k}\right)}{\sigma_k ^3 \Gamma \left(\frac{1}{2k}-\frac{3}{4}\right)}  \; x^2 \left(\sqrt{\frac{k^2 x^4}{\sigma_k ^4}+1}-\frac{k x^2}{\sigma_k ^2}\right)^{\frac{1}{k}}
\end{equation}
here $\sigma_k$ is the dispersion. When the $k$--parameter tend to $1$ again the standard Maxwellian distribution is attained.  
The CDF of Kaniadakis distribution is:
\begin{equation}
F_k(x)= 1 - \left( \frac{2^{\frac{5}{2}-\frac{1}{k}} \Gamma \left(\frac{7}{4}+\frac{1}{2 k}\right) \Gamma \left(\frac{1}{k}\right)
   \left(\frac{k x^2}{\sigma_k ^2}\right)^{\frac{3}{2}-\frac{1}{k}}}{\sqrt{\pi }} \right) \, 
   _3\tilde{F}_2\left(\frac{1}{2 k}-\frac{3}{4},\frac{k+1}{2 k},\frac{1}{2 k};\frac{k+2}{4
   k},\frac{1}{k}+1;-\frac{\sigma_k ^4}{k^2 x^4}\right)
\end{equation}
where $_3\tilde{F}_2$ is the regularized generalized hypergeometric function.

\end{appendix}

\end{document}